\documentclass[journal]{IEEEtran}

\usepackage[T1]{fontenc}

\usepackage{cite}

\usepackage{multirow}
\usepackage{graphicx}
\usepackage{amssymb}
\usepackage{amsmath}
\usepackage{booktabs}
\usepackage{array}
\newcolumntype{L}[1]{>{\raggedright\let\newline\\\arraybackslash\hspace{0pt}}m{#1}}
\newcolumntype{C}[1]{>{\centering\let\newline\\\arraybackslash\hspace{0pt}}m{#1}}
\newcolumntype{R}[1]{>{\raggedleft\let\newline\\\arraybackslash\hspace{0pt}}m{#1}}

\usepackage{lipsum}
\usepackage{changepage}
\usepackage{color,soul}
\usepackage{enumitem}   

\usepackage{pifont}

\usepackage{tikz}

\usepackage{lipsum}

\let\OLDthebibliography\thebibliography
\renewcommand\thebibliography[1]{
  \OLDthebibliography{#1}
  \setlength{\parskip}{0pt}
  \setlength{\itemsep}{0pt plus 0.3ex}
}

\usepackage{array}

\makeatletter
\newcommand{\thickhline}{%
    \noalign {\ifnum 0=`}\fi \hrule height 1pt
    \futurelet \reserved@a \@xhline
}
\newcolumntype{"}{@{\hskip\tabcolsep\vrule width 1pt\hskip\tabcolsep}}
\makeatother

\usepackage[ruled,vlined]{algorithm2e}

\usepackage{framed,multirow}

\usepackage{latexsym}

\usepackage{url}
\usepackage{xcolor}

\usepackage{hyperref}

\usepackage{bm}

\begin{document}

\title{Diffusion Tensor Estimation \\ with Transformer Neural Networks} 

\author{Davood Karimi and Ali Gholipour \\ Department of Radiology at Boston Children's Hospital, Harvard Medical School, Boston, MA, USA

\thanks{This study was supported in part by the National Institutes of Health (NIH) grants R01 EB018988, R01 NS106030, R01 EB031849, and R01 EB032366; in part by the Office of the Director of the NIH under award number S10OD0250111; in part by the National Science Foundation under award number 2123061; and in part by a Technological Innovations in Neuroscience Award from the McKnight Foundation. The content of this publication is solely the responsibility of the authors and does not necessarily represent the official views of the NIH or the McKnight Foundation.  \\ The dHCP dataset is provided by the developing Human Connectome Project, KCL-Imperial-Oxford Consortium funded by the European Research Council under the European Union Seventh Framework Programme (FP/2007-2013) / ERC Grant Agreement no. [319456]. We are grateful to the families who generously supported this trial. \\ D. Karimi and A. Gholipour are with the Computational Radiology Laboratory of the Department of Radiology at Boston Children's Hospital, and Harvard Medical School, Boston, MA, USA (email: davood.karimi@childrens.harvard.edu).}}

\maketitle

\begin{abstract}

Diffusion tensor imaging (DTI) is a widely used method for studying brain white matter development and degeneration. However, standard DTI estimation methods depend on a large number of high-quality measurements. This would require long scan times and can be particularly difficult to achieve with certain patient populations such as neonates. Here, we propose a method that can accurately estimate the diffusion tensor from only six diffusion-weighted measurements. Our method achieves this by learning to exploit the relationships between the diffusion signals and tensors in neighboring voxels. Our model is based on transformer networks, which represent the state of the art in modeling the relationship between signals in a sequence. In particular, our model consists of two such networks. The first network estimates the diffusion tensor based on the diffusion signals in a neighborhood of voxels. The second network provides more accurate tensor estimations by learning the relationships between the diffusion signals as well as the tensors estimated by the first network in neighboring voxels. Our experiments with three datasets show that our proposed method achieves highly accurate estimations of the diffusion tensor and is significantly superior to three competing methods. Estimations produced by our method with six diffusion-weighted measurements are comparable with those of standard estimation methods with 30-88 diffusion-weighted measurements. Hence, our method promises shorter scan times and more reliable assessment of brain white matter, particularly in non-cooperative patients such as neonates and infants.

\end{abstract}


\IEEEpeerreviewmaketitle

\section{Introduction}
\label{Introduction_section}

\subsection{Diffusion tensor imaging}

\IEEEPARstart{D}{iffusion} weighted magnetic resonance imaging (DW-MRI) is one of the most common medical imaging modalities. It uses the diffusion of water molecules, and restrictions thereof due to obstacles such as membranes and fibers, to generate contrast. Although its applications are not limited to brain, DW-MRI is presently the best non-invasive tool for studying the brain micro-structure in vivo. Diffusion tensor imaging (DTI) is a specific type of DW-MRI that is widely used in neuroimaging \cite{alexander2007diffusion,le2001diffusion,basser2011microstructural}. DTI is capable of capturing diffusion anisotropy, i.e., the dependence of the diffusion process on the orientation . In DTI, a Gaussian model of diffusion is assumed, whereby the orientation-dependence of diffusion is characterized with a $3 \times 3$ symmetric matrix, i.e., a tensor:

\begin{equation}
    D = \begin{bmatrix}
       D_{xx} & D_{xy} & D_{xz} \\
       D_{xy} & D_{yy} & D_{yz} \\
       D_{xz} & D_{yz} & D_{zz} \\
     \end{bmatrix}.
\end{equation}

The diffusion tensor formalism can be interpreted as representing the surface of the diffusion front with an ellipsoid. An eigen-decomposition of $D$ gives us the direction of the strongest diffusion as well as parameters such as mean diffusivity (MD) and fractional anisotropy (FA), which are widely used to study brain development and as biomarkers for various diseases \cite{lebel2008microstructural,barnea2005white,kubicki2007review,barnea2004white}. There exist more complex models of tissue micro-structure, e.g. \cite{assaf2005composite,zhang2012noddi,palombo2020sandi}. Nonetheless, DTI remains the most widely used method in brain micro-structure studies because of easier acquisition and model fitting and wide-spread availability of DTI analysis software.

Given diffusion signals measured with higher and lower diffusion weightings, $s_1$ and $s_0$ respectively, one can write:

\begin{equation}
\begin{aligned}
s_1/s_0= & \exp (-b_{xx} D_{xx} -b_{yy} D_{yy} -b_{zz} D_{zz} \\ 
& \hspace{7mm} -2 b_{xy} D_{xy} -2 b_{xz} D_{xz} -2 b_{yz} D_{yz} ),
\end{aligned}
\end{equation}

\noindent where $b{..}$ are elements of the so-called b-matrix \cite{jones2009gaussian,le2001diffusion}. Given a set of $m$ such measurements, after log-transformation, one can express the relation between the diffusion tensor elements and the measurements as a linear system of equations:

\begin{equation}
B \tilde{D}= S, 
\end{equation}

\noindent where $B$ is a design matrix that depends only on the directions and strengths of the applied diffusion-sensitizing gradients, S is the vector of log-transformed diffusion signal measurements, and $ \tilde{D}=[D_{xx}, D_{yy},$ $ D_{zz}, D_{xy}, D_{xz}, D_{yz}]$ is the vector of unknowns. 

Many different approaches have been proposed for estimating $\tilde{D}$. The ordinary least squares solution can be obtained as $\tilde{D}= (B^TB)^{-1}B^TS$. This solution is based on the assumption of homoskedasticity, i.e., that the variance of the noise is the same for all measurements. Even though this assumption is largely correct for the diffusion signal before the log transformation, it is not correct after the transformation. After log-transformation, the measurement variance is higher for lower signal intensities. Therefore, one can improve upon the ordinary least squares method by introducing weights: $\tilde{D}= (B^TWB)^{-1}B^TWS$, where $W$ is a diagonal matrix with the diagonal elements proportional to the measurements \cite{koay2006unifying}. Alternatively, one could attempt solving the original non-linear system of equations without log-transforming the measurements. This approach is theoretically more appealing but can suffer from other problems such as sensitivity to the initial solution, convergence to local minimum, and higher computational cost.

There have been attempts to improve upon the least squares-based methods mentioned above. For example, basic least squares-based methods may yield a tensor that has negative eigen-values, which is physically invalid. One approach to enforce positivity of eigenvalues is Cholesky factorization of the diffusion tensor \cite{koay2006unifying,kingsley2006introduction}. Other notable methods include algorithms that aim at reducing the effect of erroneous measurements due to such factors as high noise, subject head motion, and cardiac pulsation. For example, RESTORE algorithm uses an iterative weighted least squares strategy to detect and remove erroneous measurements \cite{chang2005restore} \cite{chang2012informed}. Bootstrap methods have been used to improve and quantify the accuracy and uncertainty of DTI parameter estimation \cite{kingsley2006introduction,chung2006comparison}.

The above methods are widely used in practice and constitute the core of the tensor fitting algorithms in common DW-MRI software \cite{tournier2019mrtrix3,garyfallidis2014dipy}. However, they require a large number of measurements for accurate tensor estimation. Although the diffusion tensor has only six degrees of freedom, practical guidelines  recommend acquiring at least 30 measurements with diffusion encoding directions uniformly spread on the sphere \cite{skare2000condition,jones1999optimal}. It is strongly recommended to increase the number of measurements to much larger than 30, if possible, in order to achieve more accurate and more robust tensor estimation \cite{jones2009gaussian}. These requirements highlight the challenging nature of estimating micro-structural parameters of interest from noisy and imperfect measurements. However, they also suggest that standard diffusion tensor estimation methods may be sub-optimal. In particular, these estimation methods are based on biophysical models of diffusion that can only approximate the true underlying signal generation processes. More importantly, the classical estimation methods fit the diffusion signal on a voxel-wise basis. They fail to take into account the correlation between signals in neighboring voxels and to exploit the spatial regularity of diffusion tensor values.

DTI estimation accuracy depends not only on the number of measurements, but also on several other factors. For example, the choice of the diffusion gradient strength (the b-vale) can have a significant effect, as shown by several prior works \cite{armitage2001utilizing,alexander2005optimal,jones1999optimal}. Similarly, the arrangement of the directions of diffusion-sensitizing gradients can also be very important \cite{jones2004effect,skare2000condition,cook2007optimal}. Another important factor is the signal to noise ratio (SNR). The effect of noise on diffusion tensor estimation error has been explored in the past and various methods for reducing the impact of measurement noise have been proposed \cite{jones2004squashing,basu2006rician,goodlett2007}. Moreover, like any other DW-MRI technique, diffusion tensor imaging suffers from artifacts such as eddy current-induced distortions, magnetic susceptibility effects, subject motion, and cardiac pulsation \cite{rohde2004comprehensive,morgan2004correction,nunes2005investigations}.

Although many factors contribute to DTI estimation error, as briefly discussed above, in general increasing the number of measurements improves the accuracy and robustness of estimation \cite{jones2004effect,jones2009gaussian}. Acquiring more measurements means longer scan times. Moreover, it may be difficult to achieve with non-cooperative subjects such as infants and young children, where part of the data may have to be discarded due to excessive motion. Therefore, methods that can accurately estimate the diffusion tensor from smaller numbers of measurements are highly desirable. Machine learning methods have a great potential in this regard. Unlike standard estimation methods, they do not need to assume a known mathematical model for the diffusion signal and noise. Instead, they learn the mapping from the diffusion signal to the tensor from training data. Furthermore, they can effectively learn the spatial correlations in the diffusion signal and the parameter(s) of interest. With the increasing availability of very large DW-MRI datasets, the advantage of machine learning methods has grown. It is now possible to train a machine learning model on these large and rich datasets and use the trained model on less perfect in-house datasets.

\subsection{Related works}

Applications of machine learning and data-driven methods for parameter estimation in DW-MRI have been explored in several prior works. Random forests, support vector regression, and other classical machine learning methods were used in several works \cite{schultz2012,neher2017fiber,nedjati2017machine,reisert2017disentangling}. More recently, deep learning has been shown to hold great promise for improving the accuracy and robustness of parameter estimation in DW-MRI \cite{golkov2016,nath2019deep,li2019fast,de2021neural}. Several recent studies have shown that deep learning can dramatically reduce the number of measurements, and hence the scan times, required for estimating micro-structural parameters of interest. The q-space deep learning (q-DL) was one of the first deep learning methods for diffusion parameter estimation \cite{golkov2016}. It showed that a three-layer neural network could accurately estimate diffusion kurtosis as well as neurite orientation dispersion and density measures, while reducing the required number of measurements by a factor of 12. This result was particularly interesting given that a very simple neural network was used and the network performed the prediction on a voxel-wise basis, i.e., without exploiting spatial correlations.

Following the success of q-DL, several studies have used deep learning models to estimate other diffusion parameters on a voxel-wise basis. One study showed that a deep neural network can achieve significantly more accurate estimation of fiber orientations than standard methods such as constrained spherical deconvolution \cite{nath2019deep}. Examples of other parameters that have been estimated on a voxel-wise basis include the number \cite{koppers2017reliable} and orientation \cite{koppers2016} of major fibers.

One can expect more accurate and more robust estimation when spatial correlations between the signal in neighboring voxels and the spatial regularity of the parameter(s) of interest are exploited. There are a variety of deep learning models that are capable of learning such spatial patterns. Perhaps the most well-known of these models are convolutional neural networks (CNN). Prior works have applied CNNs on patches of DW images to estimate the fiber orientations \cite{lin2019fast,koppers2017reconstruction}. One study reported that more accurate estimation of diffusion kurtosis measures could be obtained, compared with the q-DL framework, with a simple CNN \cite{li2019fast}. In other studies, CNNs have been used for tract segmentation and tractography analysis \cite{wasserthal2019combined,zhang2020deep}.

Several recent studies have used deep learning models for estimating the diffusion tensor, FA, and MD. One study used a CNN for direct estimation of the diffusion tensor from DW-MRI measurements \cite{li2021superdti}. Aliotta et al. used a multi-layer perceptron to estimate MD and FA \cite{aliotta2019highly}. Another notable example is the DeepDTI method \cite{tian2020deepdti}, which proposed a CNN model for estimating the diffusion signal residuals. The input to the CNN is a set of potentially noisy and artifact-full DW scans that are stacked together. The CNN learns to map these low-quality scans to their difference (residual) with respect to high-quality reference DW scans. In order to improve the model accuracy, anatomical (i.e., T1 and T2) images are registered to the DW scans and stacked with the DW scans to enrich the CNN input. In a recent study by our own team, we used CNNs for accurate color-FA estimation for fetuses \cite{karimi2021accurate,karimi2021deep}. A combination of a CNN and a multi-layer perceptron was used for FA and MD estimation in a recent work \cite{aliotta2021extracting}. Another recent study used a multi-scale encoder-decoder CNN for estimating FA and MD \cite{yigit2021quantifying}. The authors also used a monte-carlo dropout technique to compute the prediction uncertainty of the model predictions.

Despite the importance of the efforts mentioned above, the potential of deep learning for improving the accuracy and robustness of parameter estimation in DW-MRI is highly under-explored. One important shortcoming of most prior studies is their failure to effectively model the relationship between diffusion signal in neighboring voxels. Some methods, such as \cite{golkov2016,nath2019deep}, have only used the signal in one voxel. Some other methods, for example \cite{lin2019fast,koppers2017reconstruction}, have used models such as CNNs that have originally been devised for computer vision applications and are not optimal for regression and parameter estimation applications. There are also studies that have used 2D CNNs, which fail to exploit the correlations in all three dimensions \cite{gibbons2019simultaneous}.

\subsection{Contributions of this work}

In this work, we propose a novel method for diffusion tensor estimation. Our main idea is to exploit the correlations between the diffusion signal and diffusion tensor parameters in neighboring voxels. Brain tissue micro-structure is spatially regular, in the sense that micro-structural properties such as fiber orientations do not change randomly between adjacent voxels. Rather, there exist strong spatial correlations between neighboring voxels. Moreover, these spatial correlations in the brain tissue micro-structure are largely shared across the brains of different subjects. These correlations in micro-structure, in turn, give rise to correlations between diffusion signals in neighboring voxels. We propose to learn these correlations in order to improve the accuracy and robustness of DTI estimation, especially when the measurements are noisy and few in number.

Our proposed method is based on the attention models, which represent the state of the art in sequence modeling. While prior works have either ignored spatial correlations or have used computer vision models such as CNN to learn spatial correlations, we use attention models that are more flexible and more powerful. The attention mechanism has important advantages over more classical deep learning models such as CNNs and RNNs. It offers higher flexibility since network weights are adapted in an input-dependent fashion. Moreover, each component/location of the output can attend to any component/location of the input, regardless of the distance between the two components in the sequence. In RNNs, for example, it can take up to $n$ steps to propagate the information from one location in the sequence to another location, where $n$ is the sequence length. In attention models, on the other hand, information can be exchanged between any two locations in the sequence in one step. We leverage these advantages to develop a novel method for DTI estimation. We show that our proposed method can accurately estimate the diffusion tensor from only six diffusion-weighted measurements. In particular, we demonstrate the effectiveness of our method on neonatal subjects where tensor estimation is very challenging and standard estimation methods can be highly inaccurate and unreliable.

\section{Materials and methods}

\subsection{Attention models}

The concept of attention can be employed in different machine learning models, but here our focus is on neural networks. In a standard neural network, the output of each layer is computed from the output of the preceding layer using a function of the form $x^i= \Psi ({W^i}^T x^{i-1}) $, where $W^i$ represents the layer weights (for simplicity of presentation we omit the bias term) and $\Psi$ is some non-linear function. In standard neural networks, the weights are optimized on a set of training data and they remain fixed afterwards. Attention models represent an alternative paradigm in which the weights are computed dynamically based on the input. In other words, $x^i= \Psi (f_{\theta}(x^{i-1})^T  x^{i-1}) $, where $f_{\theta}$ is a learnable function \cite{murphy2012}. In this paradigm, the model weights are not fixed at training; rather, they depend on the input at inference time. The attention mechanism can take different forms. One common form is self-attention, where elements of ${W^i}$ depend on the pair-wise similarity between elements of the input sequence $x^{i-1}$. In other words: ${W^i}_{s,t}= \textbf{score} (x^{i-1}_s, x^{i-1}_t)$. The score function, $\textbf{score}(x^{i-1}_s, x^{i-1}_t)$, quantifies the similarity between the two vectors $x^{i-1}_s$ and $x^{i-1}_t$ and can take various forms. One common form is Luong's multiplicative formulation: $\textbf{score}(a, b)= a^T H b$, for some matrix $H$ \cite{luong2015effective}. Therefore, with this formulation the network can learn to compute $x^i$ by ``paying attention" to the relevant pieces of information in $x^{i-1}$ in a dynamic input-dependent manner.

In this work we follow a self-attention approach similar to that of the transformer network \cite{vaswani2017attention}. Let us denote the output of the previous network layer with $x^{i-1} \in {\rm I\!R}^{n,d}$, where $n$ is the sequence length and $d$ is the dimension of each element of the sequence. First, a set of query, key, and value sequences are computed via linear projections of $x^{i-1}$:

\begin{equation}  \label{eq:QKV}
Q^i= x^{i-1} W^i_Q , \; \; K^i= x^{i-1} W^i_K, \; \text{ and } \; V^i= x^{i-1} W^i_V
\end{equation}

The projection matrices $W^i_Q$, $W^i_K$, and $W^i_V$ are of size ${\rm I\!R}^{d,d_h}$, which means that the query, key, and value sequences will be of size $d_h$. The self-attention output is then computed as:

\begin{equation} \label{eq:self-attention}
{x^i}^*= \frac{Q^i {K^i}^T}{ \sqrt{d_h} } V^i,
\end{equation}

\noindent where the scaling factor $1/\sqrt{d_h}$ is introduced for stable computations. In other words, self-attention is formulated based on the similarity between queries and keys, which are both computed from the input $x^{i-1}$. Note that similarity between query and key vectors in Eq. \eqref{eq:self-attention} is computed as a dot product, which is a special case of Luong's attention where $H$ is the identity matrix. Since ${x^i}^*$ is in ${\rm I\!R}^{d_h}$, it is passed through another fully-connected layer to generate $x^i \in {\rm I\!R}^{d}$ as the input for the next stage of the network. A transformer network consists of a succession of such self-attention modules. The output of the last module is projected onto the space of the desired network output using a fully-connected layer.

There are two important variations to the standard transformer model that we also utilize in this work \cite{murphy2012,vaswani2017attention}. First, in order to improve the expressive power of the learned attention maps, multi-headed attention is used. In this approach, $n_h$ different query, key, and value sequences are computed, each with different projection matrices in Eq. \eqref{eq:QKV}. Then, ${x^i}^*$ is formed by concatenating the $n_h$ sequences, each computed using Eq. \eqref{eq:self-attention}. Second, the standard self-attention model lacks a means of knowing the sequence order because it is permutation-invariant. To overcome this limitation, a positional encoding is added to the input sequence \cite{vaswani2017attention}. Specifically, the sequence of initial input signals $x_s \in {\rm I\!R}^{n,d_s}$ is projected onto ${\rm I\!R}^{d}$ and a sequence of the same shape is added to it: $x^0= x_s {W_s} + p$. Here $W_s \in {\rm I\!R}^{d_s,d}$ is the signal embedding projection matrix and $p\in {\rm I\!R}^{n, d}$ is meant to encode the relative position between elements of the input sequence. Many different forms of positional encoding have been proposed \cite{shiv2019novel,vaswani2017attention,dosovitskiy2020image,murphy2012}. In this work, because we do not know \textit{a priori} how diffusion signals and tensors in neighboring voxels are related, we consider a learnable positional encoding. In other words, in our method $p$ is a free parameter that is learned during training.

\subsection{Proposed method}
\label{proposed_method}

\begin{figure*}
\centering
\includegraphics[width=\textwidth]{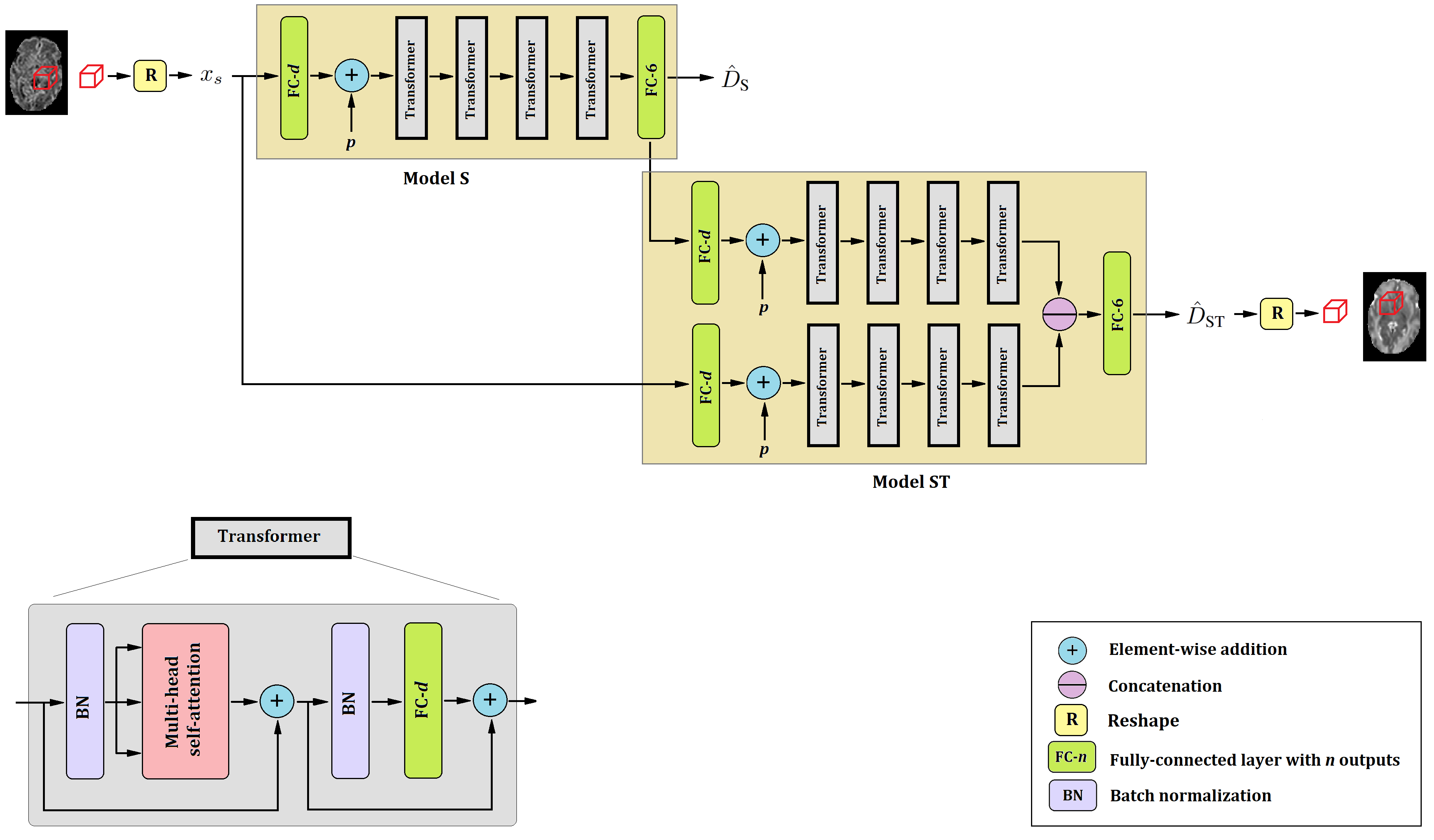}
\caption{A schematic of the proposed method. The input image is a volumetric (3D) image with six channels, where each channel represents one of the six normalized diffusion-weighted measurements. The output is a tensor image of the same shape as the input image with six channels, where each of the channels represents one of the six tensor elements.} 
\label{method}
\end{figure*}

Figure \ref{method} shows our proposed method, which we refer to as Transformer-DTI because its core is similar to the transformer model \cite{vaswani2017attention,murphy2012}. It requires only six diffusion weighted measurements. The directions of the diffusion-sensitizing gradients for these six diffusion weighted measurements are clarified in more detail below. To exploit the spatial correlations in the signal and tissue micro-structure, the model uses the signal in a cubic patch of side length equal to $L$ voxels to estimate the diffusion tensor in a cubic patch of the same size. Unless otherwise specified, in the experiments reported in this paper we use $L=5$.

The proposed model aims to estimate the diffusion tensor in each voxel by exploiting the spatial correlations in the signal and in the diffusion tensor. To this end, our model consists of two sub-models, which are trained separately and sequentially. Model S uses the diffusion signal as the input and estimates the diffusion tensor. Model ST, on the other hand, uses the diffusion signal as well as the diffusion tensor estimated by Model S to provide a more accurate estimation of the diffusion tensor. As we show in the Results section below, Model S on its own can provide accurate tensor estimations. Nonetheless, the final estimations provided by Model ST are significantly more accurate. This is because Model S can only exploit the correlations between diffusion signals in the neighboring voxels. Model ST builds upon the estimation produced by Model S. Although the tensors estimated by Model S are not optimal, they serve as useful additional input for Model ST. Therefore, Model ST which provides our final tensor estimate, can learn to exploit the correlations in the diffusion signal as well as the correlations in the diffusion tensor.

Denoting the number of diffusion measurements in each voxel with $n_s$, the diffusion signal in a patch will be of shape ${\rm I\!R}^{L,L,L,n_s}$. This is first reshaped into $x_s \in {\rm I\!R}^{n,n_s}$, where $n=L^3$, which will serve as the input signal sequence for both Model S and Model ST. In both models, $x_s$ is first embedded into ${\rm I\!R}^{d}$ and a learnable positional encoding sequence is added to form the input to the transformer modules, $x_0 \in {\rm I\!R}^{n,d}$. In Model S, this position-encoded sequence is passed through a series of $N_{\text{Tr}}$ transformer modules. The output of the last transformer module is projected from ${\rm I\!R}^{d}$ onto ${\rm I\!R}^{6}$ to form the estimated tensor sequence of size ${\rm I\!R}^{L^3,6}$. This output sequence can be reshaped into ${\rm I\!R}^{L,L,L,6}$ to obtain the tensor estimate for the input patch.

Model ST has two branches, each one of them similar to Model S. One of the branches works on the diffusion signal, similar to Model S. The other branch is architecturally identical, but works on the diffusion tensor estimated by Model S. These two branches are meant to learn the spatial correlations between the diffusion signal and the tensor values. The outputs of these two branches are concatenated and passed through a fully-connected layer to estimate the final diffusion tensor estimate for the patch.

\subsection{Implementation and training}
\label{implementation}

We selected the model architecture hyper-parameters using preliminary experiments on our training datasets. We set the number of transformer modules, $N_{\text{Tr}}$, in Model S and each of the two branches of Model ST to 4, as shown in Figure \ref{method}. Furthermore, we set $d_h=512$, and the number of heads in multi-headed self-attention $n_h=2$. We discuss the effects of some of these hyper-parameters on the performance of the proposed method in the Results section below. We initialized all learnable parameters using He's method \cite{he2015}. We first trained Model S, by minimizing the square of the difference between estimated ($\hat{D}_{\text{S}}$) and reference ($D_{\text{ref}}$) tensors:

\begin{equation} \label{eq:loss}
\mathcal{L} ( \hat{D}_{\text{S}}, D_{\text{ref}} ) = \| \hat{D}_{\text{S}} - D_{\text{ref}} \|_2^2 
\end{equation}

Once training of Model S was finished, we trained Model ST. Our experiments showed that further fine-tuning Model S during the training of Model ST did not improve the accuracy of Model ST. Therefore, while training Model ST, we kept Model S fixed. Model ST was also trained using a loss function similar to Eq. \ref{eq:loss}, with $\hat{D}_{\text{ST}}$ in place of $\hat{D}_{\text{S}}$. Both models were trained with a batch size of 10, and an initial learning rate of $10^{-4}$ using Adam \cite{kingma2014}. We reduced the learning rate by a factor of 0.9 every time the loss on the validation set did not decrease after a training epoch. We stopped the training if the validation loss did not decrease after two consecutive epochs. All models were implemented in TensorFlow. Training and validation were performed on a Linux computer with an NVIDIA GeForce GTX 1080 GPU.

\subsection{Data and evaluation approach}
\label{validation}

Most of the experimental results presented in this paper are with the developing Human Connectome Project (dHCP) dataset \cite{bastiani2019}. Nonetheless, to demonstrate the applicability of our method to other datasets, we also report experimental results with scans from the Pediatric Imaging, Neurocognition, and Genetics (PING) dataset \cite{jernigan2016pediatric} as well as in-house scans of Vein of Galen Malformation (VOGM) patients at Boston Children's Hospital. This study was approved by the institutional review board. We mostly focus on the dHCP dataset because it is a publicly-available dataset on which the interested reader can train and test our method and because it is a challenging dataset. It consists of DW-MRI images of neonates scanned at 29-45 gestational weeks. The neonatal period represents a critical time in brain development. It is characterized by rapid cortical expansion and formation of connections between distant regions of the brain. Immature myelination of the white matter and patient motion make diffusion tensor imaging of neonates especially challenging.

Each DW-MRI scan in the dHCP dataset includes measurements at three b-values of 400, 1000, and 2600. Following the widely-adopted recommendations \cite{jones1999optimal}, we use the $b=1000$ measurements for DTI estimation. Each scan includes 88 measurements in the $b=1000$ shell, which are approximately uniformly distributed on the sphere. For each subject, we used all 88 measurements to reconstruct a high-quality ``reference" DTI with the constrained weighted linear least squares (CWLLS) method \cite{koay2006unifying}. We refer to this reference reconstruction as $D_{\text{ref}}$. For our proposed method, we selected six of the 88 measurements that were closest to the six optimized directions proposed in \cite{skare2000condition}. Specifically, the unit vectors indicating these directions are: $[0.910, \pm 0.416,$ $ 0.000]$, $[\pm 0.416, 0.000, 0.910]$, and $[0.000, 0.910, \pm 0.416]$. These six directions have been derived to minimize the condition number of the diffusion tensor transformation matrix \cite{skare2000condition}. We also selected one of the $b=0$ measurements to normalize the six $b=1000$ measurements. We refer to the reconstructions of our proposed method with these six normalized measurements as $D_{\text{Tr}}$. For comparison with existing methods, we apply three methods on the same six diffusion-weighted measurements and one $b=0$ measurement. These three methods are the following: 1) Constrained Weighted Linear Least Squares (CWLLS) \cite{koay2006unifying}, which is the standard estimation method, 2) Constrained Nonlinear Least Squares (CNLS) \cite{koay2006unifying}, and 3) The CNN-based method of Lin et al. \cite{lin2019fast}. This method was originally proposed for estimation of fiber orientation distribution. We simply changed the first and the last layers of the network to match our application. We refer to this method as CNN-DTI.

With the dHCP dataset, we trained our method using scans of 200 subjects aged 40-45 gestational weeks. We used data from 40 of these subjects as validation set during various stages of hyper-parameter selection and training. Once the training of the final model was complete, we tested our method on scans of 40 independent subjects; 20 of these were from the same age range, while the other 20 were younger subjects aged 29-36 gestational weeks. We compare our method with the competing techniques in terms of the norm of the difference between the estimated tensor and the reference tensor, $D_{\text{ref}}$. Specifically, if we denote the predicted tensor with $D_{\text{pred}}$, then we define the tensor estimation error as $\sum_{i=1}^6 | D_{\text{pred}}^i - D_{\text{ref}}^i |$, where the index $i$ refers to the six tensor elements. Moreover, we compute the error in FA, MD, and the angle of the major eigen-vector of the tensor. To do this, we calculate the eigen-decomposition of the tensor in each voxel and compute the FA and MD from the eigen-values using their standard definitions \cite{jones2009gaussian}. For each voxel, the error in FA and MD is defined as the absolute difference between those of the predicted and reference tensors. We compute the angle between the major eigen-vectors of the predicted and reference tensors as the angular error between the two tensors. We also present tractography and connectivity analysis results. To have a fair comparison between different methods, no especial data pre-processing (e.g., smoothing or re-sampling) was performed for our method or any of the compared techniques. Our proposed method does not rely on such data pre-processing steps.

\begin{table*}[!htb]
\footnotesize
 \caption{Estimation error on older (40-45 gestational weeks) and younger (29-36 gestational weeks) test subjects from the dHCP dataset. Bold type indicates statistically smaller errors at $p=0.001$, computed using paired t-tests to compare our proposed method with every one of the competing methods}.
\label{table:results_dhcp}
\begin{tabular}{ L{35mm}  L{14mm} C{30mm} C{30mm} C{16mm} C{19mm}  }
\thickhline
Test subjects & method & tensor ($\times 1000$), $mm^2s^{-1}$ & MD ($\times 1000$), $mm^2s^{-1}$ & FA & angle (degrees)  \\ \thickhline
\multirow{4}{*}{Older neonates ($n=20$)} & CWLLS &  $0.450 \pm 0.040$ & $0.413 \pm 0.042$ & $0.155 \pm 0.017$ & $20.2 \pm 2.45$  \\
& CNLS &     $0.438 \pm 0.044$ & $0.410 \pm 0.041$ & $0.149 \pm 0.019$ & $20.3 \pm 2.48$  \\
& CNN-DTI &  $0.393 \pm 0.037$ & $0.345 \pm 0.037$ & $0.124 \pm 0.011$ & $17.2 \pm 2.20$  \\
& Proposed & $\bm{0.118 \pm 0.019}$ & $\bm{0.042 \pm 0.029}$ & $\bm{0.071 \pm 0.003}$ & $\bm{11.6 \pm 2.07}$  \\ \hline
\multirow{4}{*}{Younger neonates ($n=20$)} & CWLLS &  $0.430 \pm 0.058$ & $0.376 \pm 0.050$ & $0.141 \pm 0.019$ & $18.5 \pm 2.85$  \\
& CNLS &     $0.420 \pm 0.060$ & $0.370 \pm 0.050$ & $0.144 \pm 0.022$ & $18.7 \pm 2.87$  \\
& CNN-DTI &  $0.401 \pm 0.049$ & $0.325 \pm 0.041$ & $0.137 \pm 0.017$ & $16.4 \pm 2.34$  \\
& Proposed & $\bm{0.122 \pm 0.018}$ & $\bm{0.123 \pm 0.022}$ & $\bm{0.073 \pm 0.012}$ & $\bm{10.3 \pm 2.75}$  \\
\thickhline
\end{tabular}
\end{table*}

\section{Results and Discussion}

Table \ref{table:results_dhcp} shows the error in the estimated tensor and three tensor-derived variables, i.e., FA, MD, and the orientation of the major eigen-vector for the proposed method and the three compared methods. For each of these variables, we computed the average error (across all brain voxels) separately for each test subject, thereby obtaining one error value for each parameter and subject. The numbers in Table  \ref{table:results_dhcp} and the other tables in this paper show the mean $\pm$ standard deviation across subjects. On all 40 test subjects from the dHCP datasets our method achieved lower estimation errors than all other methods. As shown in the table, compared with other methods, our proposed method has reduced the error in MD by factors of approximately 2.6-9.8 and the error in FA and orientation of major eigen-vector by factors of approximately 1.5-2.2. We ran paired t-tests to determine the statistical significance of the differences between our method and the other methods. These tests showed that the errors for the proposed method were significantly lower ($p<0.001$) than those of the three compared methods.

\begin{figure*}[!htb]
\centering
\includegraphics[width=\textwidth]{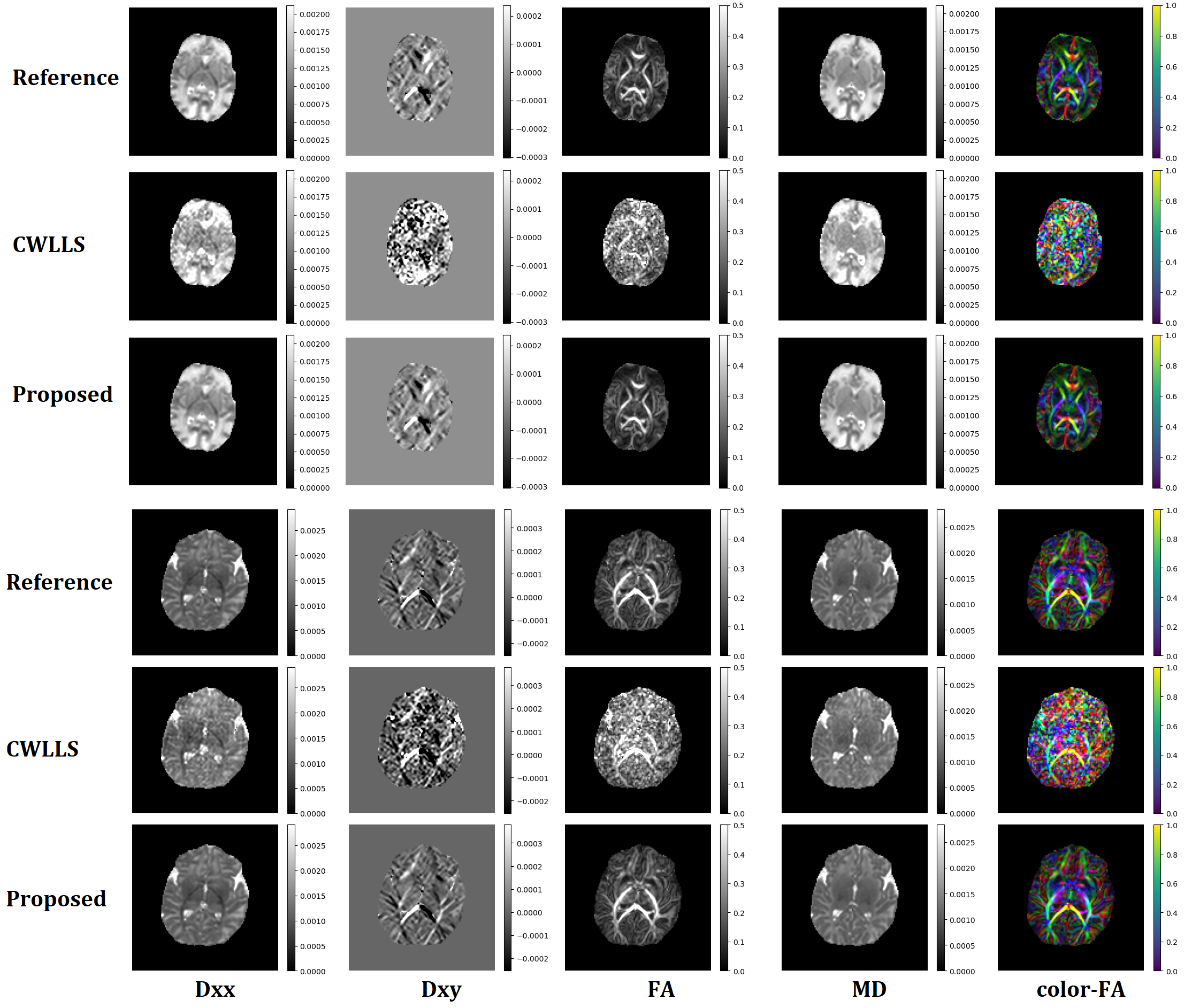}
\caption{Example tensor images estimated with CWLLS and the proposed method and their corresponding tensor-derived parameters. These images were reconstructed from scans of two neonatal subjects from the dHCP dataset. The top subject has a gestational age of 41 weeks, which is within the age range of subjects that have been used for training (i.e., 40-45 gestational weeks). The bottom subject has a gestational age of 31 weeks, which is much younger than the age range of the training subjects. For each subject, we have shown two of the six tensor channels (i.e., $D_{xx}$ and $D_{xy}$), FA, MD, and color-FA.}
\label{figure_dhcp}
\end{figure*}

Figure \ref{figure_dhcp} shows examples images reconstructed with the proposed method and CWLLS for two test subjects from the dHCP dataset. It shows two of the tensor channels, FA, MD, and color-FA images. Due to space limits, in this figure and the following figures we show the results of selected compared methods. The results shown in Figure \ref{figure_dhcp} clearly demonstrate a substantial advantage for the proposed method compared with CWLLS. The parameters estimated with the proposed method using six diffusion-weighted measurements are close to the reference images obtained with 88 measurements. On the other hand, CWLLS estimations are very noisy and contain large errors both in the gray matter area as well as in the location of major white matter tracts. Visually, the superiority of our method compared with CWLLS can be best seen in the color-FA images. These are standard color-coded FA images that display the tensor anisotropy and the orientation of the major eigenvector in a single image. These images show that our method can accurately estimate the diffusion tensor throughout the brain.

\begin{figure*}[!htb]
\centering
\includegraphics[width=172mm]{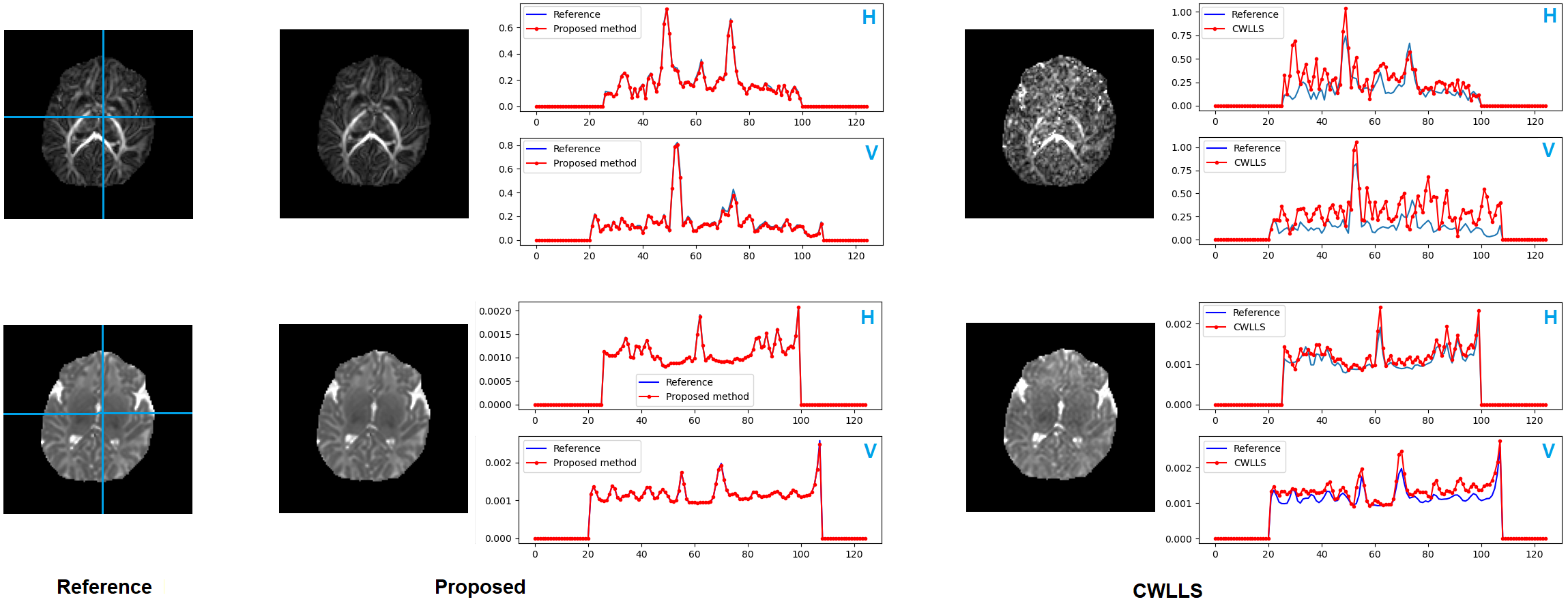}
\caption{Example one-dimensional profiles of FA (top) and MD (bottom) images reconstructed by the proposed method and CWLLS. For each image, we have shown one horizontal (H) and one vertical (V) profile. The locations of these profiles have been marked with the blue lines on the reference images (left-most column).}
\label{figure_profiles}
\end{figure*}

\begin{figure*}[!htb]
\centering
\includegraphics[width=172mm]{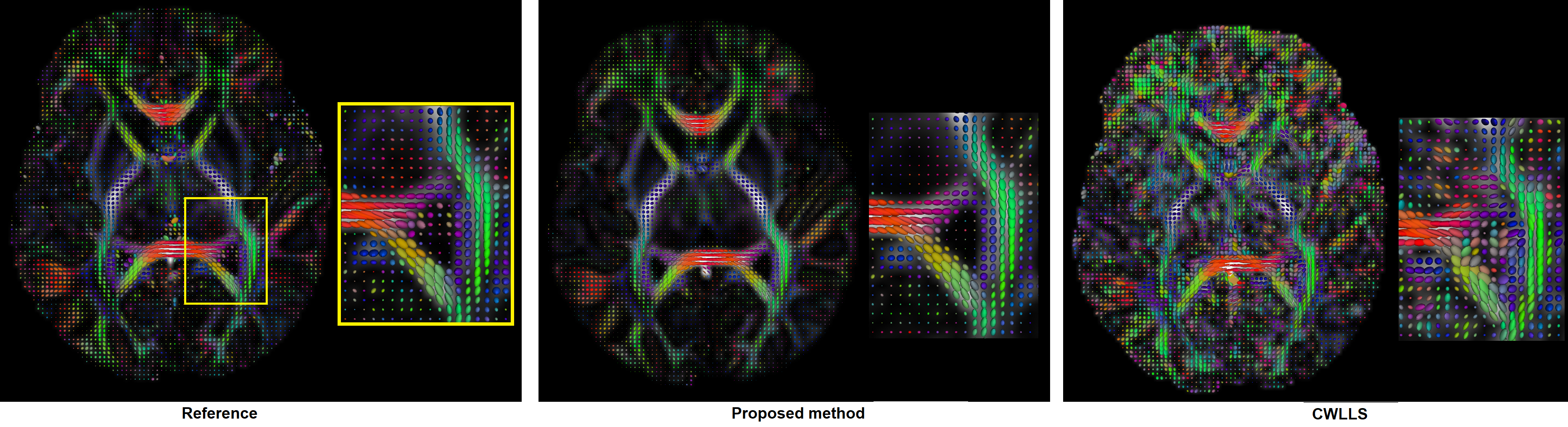}
\caption{Example glyph visualizations of the diffusion tensor images reconstructed with the proposed method and CWLLS alongside the reference tensors.}
\label{figure_glyphs}
\end{figure*}

From the examples shown in Figure \ref{figure_dhcp}, the images reconstructed with our method seem to be less noisy than the reference image. Although the reference image is based on 88 measurements, it is computed by fitting the diffusion tensor on a voxel-wise basis, i.e., by considering the diffusion signal in each voxel, one at a time. Our proposed method, on the other hand, uses the correlations between the diffusion signal and tensor values among $L^3$ neighboring voxels. For further visual assessment of the reconstruction results, in Figure \ref{figure_profiles} we have shown example one-dimensional profiles from FA and MD images reconstructed with our method and CWLLS compared with the reference. Furthermore, in Figure \ref{figure_glyphs} we have shown glyph visualization of the tensors. The profiles in Figure \ref{figure_profiles} show that, compared with the reference, the reconstructions produced by our method show no significance loss of edge sharpness and are close to the reference image in almost all locations. The reconstructions produced with CWLLS are noisy and very different from the reference. The tensor visualizations presented in Figure \ref{figure_glyphs} further show that our method is close to the reference for most brain regions. CWLLS reconstructions, on the other hand, show large estimation errors in this figure.

\begin{figure*}[!htb]
\centering
\includegraphics[width=170mm]{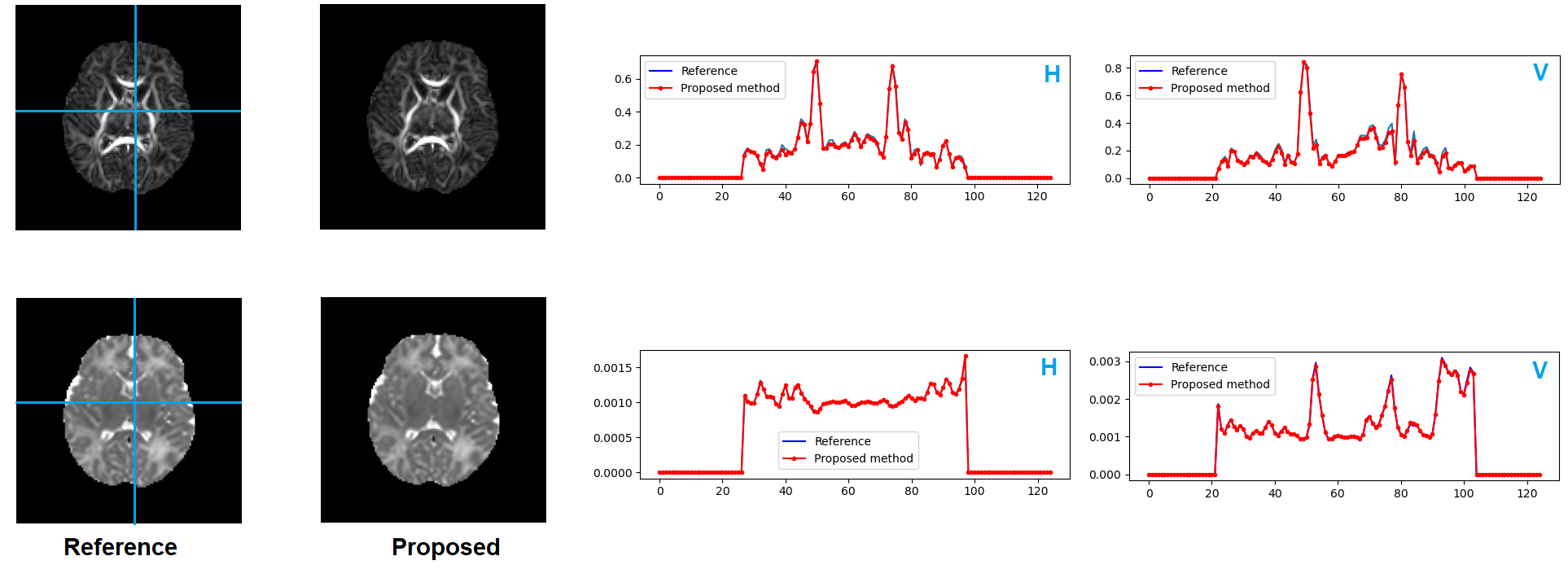}
\caption{Example slices (left) and one-dimensional profiles (right) from the FA and MD images reconstructed with the proposed method with added simulated Rician noise with an SNR of 15dB. For each image, we have shown one horizontal (H) and one vertical (V) profile. The locations of these profiles have been marked with the blue lines on the corresponding reference image slice (left-most column).}
\label{figure_snr}
\end{figure*}

In order to investigate the effect of measurement noise on the performance of our proposed method, we conducted a simulation experiment where we added simulated noise to the scans from the dHCP dataset. Specifically, independent and identically-distributed Rician noise with signal-to-noise ratio (SNR) in the range [15, 50] dB was added. An SNR of 15 dB is close to the lowest SNR reported or used for simulation in prior works (\cite{tournier2007}, \cite{dell2010modified}). Table \ref{table:table_noise} shows the results of this experiment in terms of error in FA and the orientation of the major tensor eigen-vector for the proposed method and CWLLS. The noise had a much smaller effect on our proposed method than on CWLLS. For our method, the error in FA and the orientation of the major eigenvector increased by 4.2\% and 5.2\%, respectively, as we reduced the SNR from 50dB to 15dB. Comparatively, the errors for CWLLS increased by 18.1\% and 20.0\%, respectively. Figure \ref{figure_snr} shows example slices and profiles of FA and MD images reconstructed with our proposed method at SNR=15. They show that, even at this low SNR, reconstructions of our proposed method are close to the reference image.

\begin{table*}[!htb]
\footnotesize
 \caption{Reconstruction error in terms of the angle of the major tensor eigen-vector and FA for CWLLS and the proposed method for different amounts of simulated noise added to the signal. The test subjects in this experiment were the 20 older neonates that were used in the experiment reported in Table \ref{table:results_dhcp}. The first column shows the case where no noise was added to the signal. Bold type indicates statistically smaller errors at $p=0.001$, computed using paired t-tests to compare our proposed method with CWLLS.}
\label{table:table_noise}
\begin{tabular}{ L{25mm}  L{15mm} C{20mm} C{20mm} C{20mm} C{20mm} C{20mm}  }
\thickhline
Parameter & method & No added noise & SNR= 50dB & SNR= 30dB & SNR= 20dB & SNR= 15dB  \\ \thickhline
\multirow{2}{*}{ angle (degrees) } & CWLLS & $20.2 \pm 2.45$ &  $20.5 \pm 2.41$  & $21.2 \pm 2.49$  & $22.8 \pm 2.55$ & $24.6 \pm 2.56$ \\
& Proposed & $\bm{11.6 \pm 2.07}$ & $\bm{11.7 \pm 2.09}$ & $\bm{11.9 \pm 2.05}$ & $\bm{12.0 \pm 2.11}$ & $\bm{12.2 \pm 2.15}$  \\ \hline
\multirow{2}{*}{ FA } & CWLLS &  $0.155 \pm 0.017$  & $0.157 \pm 0.017$ & $0.165 \pm 0.020$ & $0.172 \pm 0.025$ & $0.183 \pm 0.026$ \\
& Proposed & $\bm{0.071 \pm 0.003}$ & $\bm{0.071 \pm 0.003}$ & $\bm{0.071 \pm 0.005}$ & $0.073 \pm 0.006$ & $\bm{0.074 \pm 0.006}$  \\
\thickhline
\end{tabular}
\end{table*}

\begin{figure}[!htb]
\centering
\includegraphics[width=8.8cm]{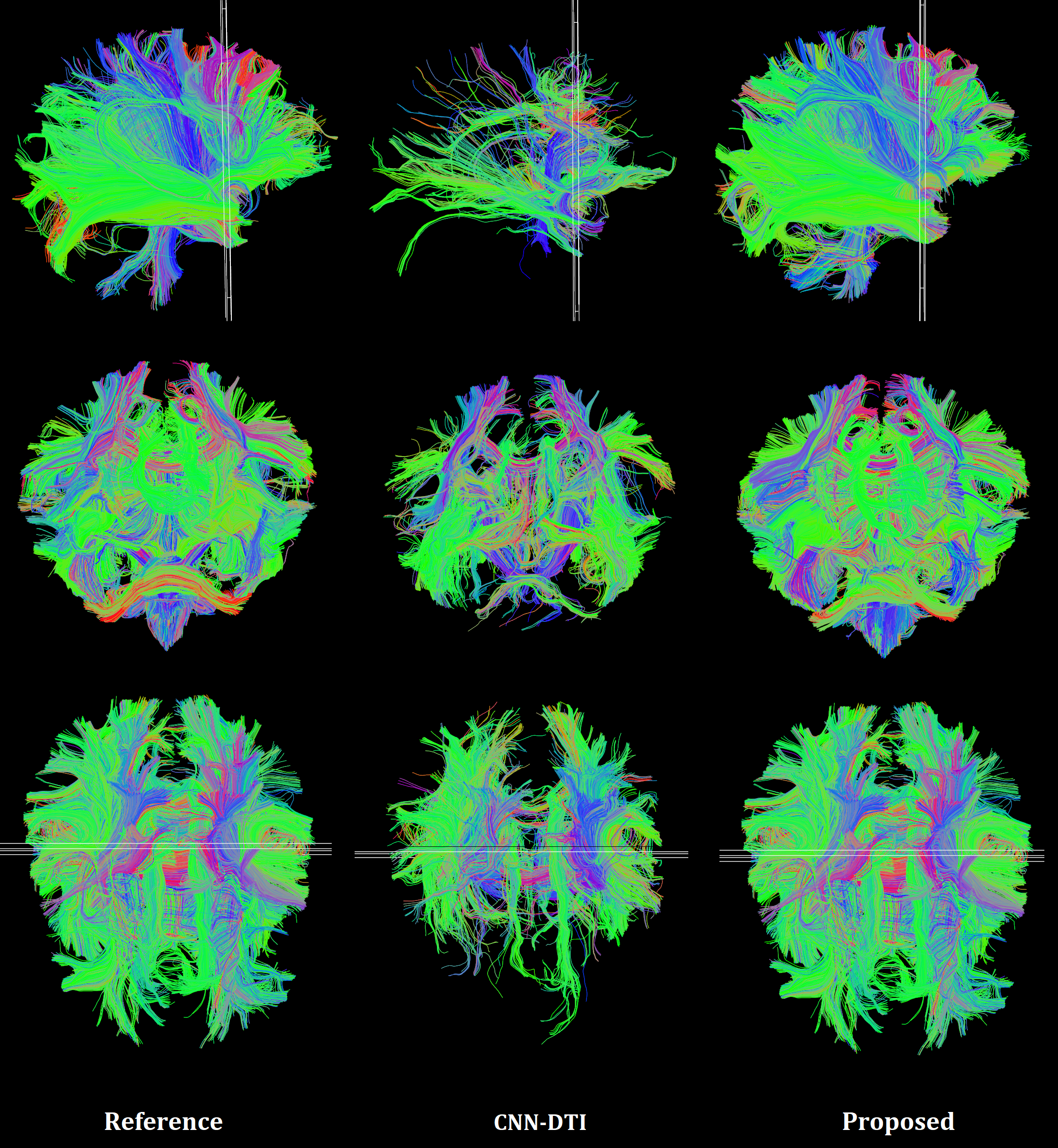}
\caption{Example whole-brain connectomes produced with the EuDX tractography algorithm based on the reference diffusion tensor estimated from 88 measurements and the diffusion tensors estimated with our proposed method and CNN-DTI from only 6 measurements.}
\label{figure_trk}
\end{figure}

\begin{figure}[!htb]
\centering
\includegraphics[width=8.8cm]{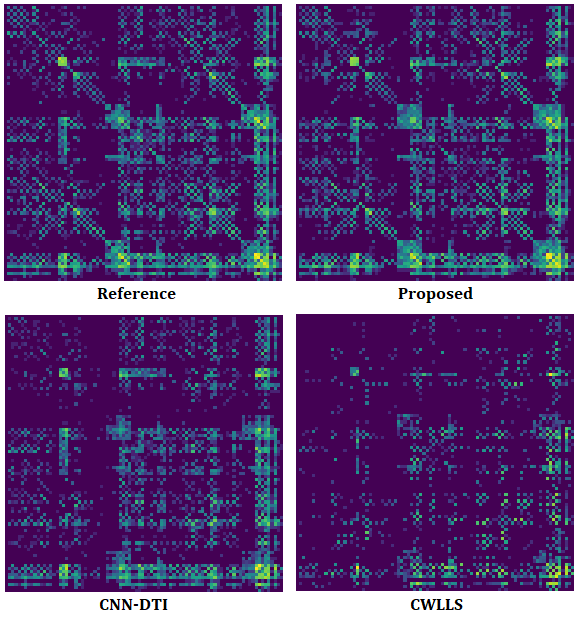}
\caption{The connectivity map between 87 brain regions for a test subject in the dHCP dataset. The connectivity maps were computed from the whole-brain connectomes derived based on the reference diffusion tensor (estimated from 88 measurements) and the diffusion tensors estimated with the proposed method, CWLLS, and CNN-DTI (estimated from 6 measurements).}
\label{connectivity_maps}
\end{figure}

Figure \ref{figure_trk} shows three example whole-brain connectomes generated from the diffusion tensors estimated with the proposed method and CNN-DTI. Tractography provides an indirect assessment of diffusion tensor estimation accuracy because the tractography results depend on the settings of the fiber tracing algorithm. For a fair comparison, we used the same seed locations and the same fiber tracing algorithm for different methods. Specifically, we used the white matter mask provided as part of the dHCP dataset for seeding. Moreover, we used the EuDX tractography algorithm \cite{garyfallidis2013towards}, which is a fiber tracing algorithm that relies heavily on voxel-wise fiber directions instead of imposing global fiber priors. We used a step size of 0.5 mm. As shown in these example figures, the connectome produced with our proposed method from only 6 measurements is similar to the one produced based on $D_{\text{ref}}$ reconstructed from 88 measurements. The connectome produced with CNN-DTI, on the other hand, lacks much of the major white matter tracts. Figure \ref{connectivity_maps} shows example connectivity maps between 87 brain regions provided for each dHCP scan, computed from the whole-brain connectomes. As shown in this example, the connectivity map for the proposed method from only 6 measurements is similar to that of the reference connectome computed from 88 measurements. The connectivity maps for CNN-DTI and CWLLS, on the other hand, are very different and lack many of the connections that are present in the reference connectivity map. For a quantitative comparison of the connectivity matrices, we considered 100 off-diagonal elements of the reference connectivity matrix with the largest values, i.e., the 100 pairs of brain regions with the strongest connections. We compared the values of those matrix elements between the reference connectivity matrix and the connectivity matrix computed with our method using a paired t-test with a significance threshold of $p=0.001$. The test showed that the connectivity matrix for our method was not different from the Reference ($p=0.32$). On the other hand, the comparison of the Reference connectivity matrix and the connectivity matrix estimated with CWLLS using the same statistical significance test showed a significant difference ($p < 0.001$). Similarly, for CNN-DTI the difference with the Reference connectivity matrix was significant ($p < 0.001$).

\begin{figure*}[!htb]
\centering
\includegraphics[width=\textwidth]{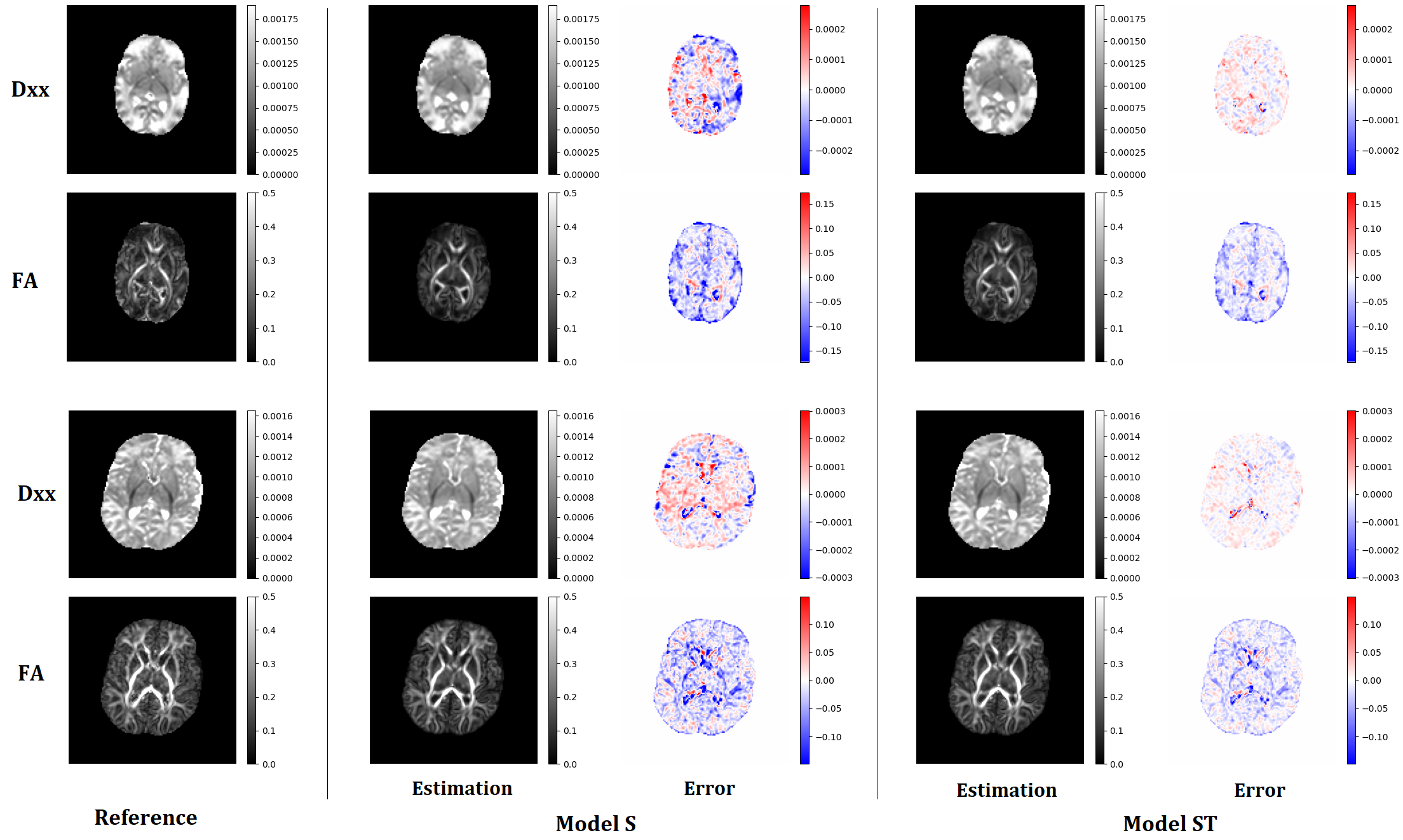}
\caption{Comparison of estimation errors for Model S and Model ST in our proposed method. We have shown the results on two subjects from the dHCP dataset; an older subject (top) and a younger subject (bottom). As shown in Figure \ref{method}, Model S uses the diffusion signals to predict the tensor values. Model ST, on the other hand, builds upon the estimations of Model S by learning to incorporate the spatial correlations in the diffusion signal and tensor values.} \label{figure_submodels}
\end{figure*}

\begin{table*}[!htb]
\footnotesize
 \caption{Average and standard deviation of the estimation error for CWLLS and the proposed method on scans of 20 subjects from the PING dataset and 7 VOGM patients. Bold type indicates statistically lower errors at $p=0.001$ computed using paired t-tests to compare our proposed method with every one of the competing methods}.
\label{table:results_ping_vogm}
\begin{tabular}{ L{23mm}  L{14mm} C{30mm} C{28mm} C{17mm} C{20mm}  }
\thickhline
Test subjects & method & tensor ($\times 1000$), $mm^2s^{-1}$ & MD ($\times 1000$), $mm^2s^{-1}$ & FA & angle (degrees)  \\ \thickhline
\multirow{4}{*}{PING ($n=20$)} & CWLLS &  $0.320 \pm 0.035$ & $0.317 \pm 0.040$ & $0.107 \pm 0.022$ & $18.5 \pm 3.30$  \\
& CNLS &     $0.322 \pm 0.037$ & $0.319 \pm 0.040$ & $0.110 \pm 0.025$ & $18.9 \pm 3.30$  \\
& CNN-DTI &  $0.299 \pm 0.030$ & $0.277 \pm 0.032$ & $0.103 \pm 0.020$ & $17.8 \pm 3.07$  \\
& Proposed & $\bm{0.174 \pm 0.021}$ & $\bm{0.101 \pm 0.033}$ & $\bm{0.073 \pm 0.011}$ & $\bm{12.4 \pm 2.28}$  \\ \hline
\multirow{4}{*}{VOGM ($n=7$)} & CWLLS &  $0.407 \pm 0.054$ & $0.350 \pm 0.049$ & $0.125 \pm 0.016$ & $18.0 \pm 2.53$  \\
& CNLS    &  $0.401 \pm 0.055$ & $0.364 \pm 0.050$ & $0.127 \pm 0.017$ & $18.2 \pm 2.77$  \\
& CNN-DTI &  $0.346 \pm 0.034$ & $0.329 \pm 0.042$ & $0.114 \pm 0.012$ & $15.9 \pm 2.22$  \\
& Proposed & $\bm{0.147 \pm 0.031}$ & $\bm{0.176 \pm 0.028}$ & $\bm{0.082 \pm 0.014}$ & $\bm{11.5 \pm 2.69}$  \\
\thickhline
\end{tabular}
\end{table*}

With regard to architectural hyper-parameters, increasing the patch size ($L$) to 7 did not improve the estimation accuracy of the proposed method. Note that the number of signals in the sequence grows cubically with $L$. Increasing $L$ from 5 to 7 would increase the sequence length from 125 to 343. Our experiments show that the transformer modules become difficult to train with $L \geq 6$. On the other hand, reducing $L$ to 3 significantly increased the method's estimation error, obviously because of the reduced spatial context to utilize in the estimation. We also found that increasing the number of transformer modules in Models S and ST beyond 4 did not improve the accuracy of our method. Furthermore, increasing the projection dimension and the number of self-attention heads to values larger than our default values (i.e., $d_h=512$ and $n_h=2$) did not improve the estimation accuracy. On the other hand, our two-stage estimation approach proved to be effective. 

Figure \ref{figure_submodels} shows example tensor and FA images reconstructed with Model S and Model ST and the corresponding error maps, i.e., the difference between the estimated values and the reference. Model ST consistently improved the estimation accuracy of Model S. For example, the tensor estimation error for Model S on older neonates and younger neonates from the dHCP dataset were, respectively, $0.154 \pm 0.028$ and $0.168 \pm 0.031$. Paired t-tests showed that these errors were significantly higher than the errors for Model ST reported in Table \ref{table:results_dhcp}, although they were significantly lower than the errors for CWLLS, CNLS, and CNN-DTI ($p<0.001$). We performed extensive experiments to investigate whether the accuracy of Model S could be improved to match Model ST by changing hyper-parameters ($L$, $d_h$, $n_h$, number of transformer modules, and training settings). However, Model ST was consistently more accurate than Model S, regardless of hyper-parameter settings. These observations support and justify our proposed two-stage approach that exploits the spatial correlations in the diffusion signal as well as in the tensors.

The error maps shown in Figure \ref{figure_submodels} display some spatial structure for both Model S and Model ST. To examine these spatial structures, we computed FA and MD estimation errors separately for white matter (WM), gray matter (GM) and CSF voxels. Furthermore, we computed these errors for four different structures in WM. The results of this analysis are presented in Table \ref{table:table_detailed_error}. The first observation from this table is that, for both FA and MD, the reconstruction errors in WM are less than those in GM and CSF. However, for the four different WM sub-structures the errors are very similar. We performed paired t-tests to compare the FA and MD reconstruction errors for these four structures, separately for Model S and Model ST. None of these tests showed statistical significance at $p=0.001$. On the other hand, on all three tissue types (WM, GM, CSF) and all four WM structures shown in this table, the FA and MD estimation errors for Model ST were significantly smaller ($p<0.001$) than those for Model S.

\begin{table*}[!htb]
\footnotesize
 \caption{FA and MD reconstruction errors for Model S and Model ST for selected tissue types: white matter (WM), gray matter (GM), and Cerebrospinal Fluid (CSF). Additionally, errors have been shown for four white matter structures: corpus callosum (CC), temporal lobe white matter (TLWM), occipital lobe white matter (OLWM), and frontal lobe white matter (FLWM). The test subjects in this experiment were 20 older neonates from the dHCP dataset. Bold type indicates statistically smaller errors at $p=0.001$, computed using paired t-tests. As in the other tables, to simplify the presentation we have multiplied the MD errors by 1000 and they are in units of $mm^2s^{-1}$.}
\label{table:table_detailed_error}
\begin{tabular}{ L{7mm}  L{13mm} | C{17mm} C{16mm} C{16mm} | C{16mm} C{16mm} C{16mm} C{16mm} C{16mm}  |  }
\thickhline
& method & WM & GM & CSF & CC & TLWM & OLWM & FLWM   \\ \thickhline
\multirow{2}{*}{ FA } & Model S & $0.058 \pm 0.004$ & $0.066 \pm 0.006$ & $0.092 \pm 0.007$ & $0.060 \pm 0.004$ & $0.058 \pm 0.005$ & $0.057 \pm 0.004$ & $0.057 \pm 0.006$   \\
& Model ST & $\bm{0.054 \pm 0.004}$ & $\bm{0.061 \pm 0.005}$ & $\bm{0.082 \pm 0.007}$ & $\bm{0.053 \pm 0.006}$ & $\bm{0.055 \pm 0.004}$ & $\bm{0.054 \pm 0.005}$ & $\bm{0.052 \pm 0.005}$   \\ \hline
\multirow{2}{*}{ MD } & Model S & $0.038 \pm 0.020$ & $0.045 \pm 0.018$ & $0.058 \pm 0.015$ & $0.037 \pm 0.022$ & $0.037 \pm 0.024$ & $0.040 \pm 0.020$ & $0.040 \pm 0.023$   \\
& Model ST & $\bm{0.035 \pm 0.018}$ & $\bm{0.041 \pm 0.017}$ & $\bm{0.050 \pm 0.016}$ & $\bm{0.034 \pm 0.020}$ & $\bm{0.034 \pm 0.021}$ & $\bm{0.036 \pm 0.017}$ & $\bm{0.035 \pm 0.015}$   \\ 
\thickhline
\end{tabular}
\end{table*}

\begin{figure*}[!ht]
\centering
\includegraphics[width=172mm]{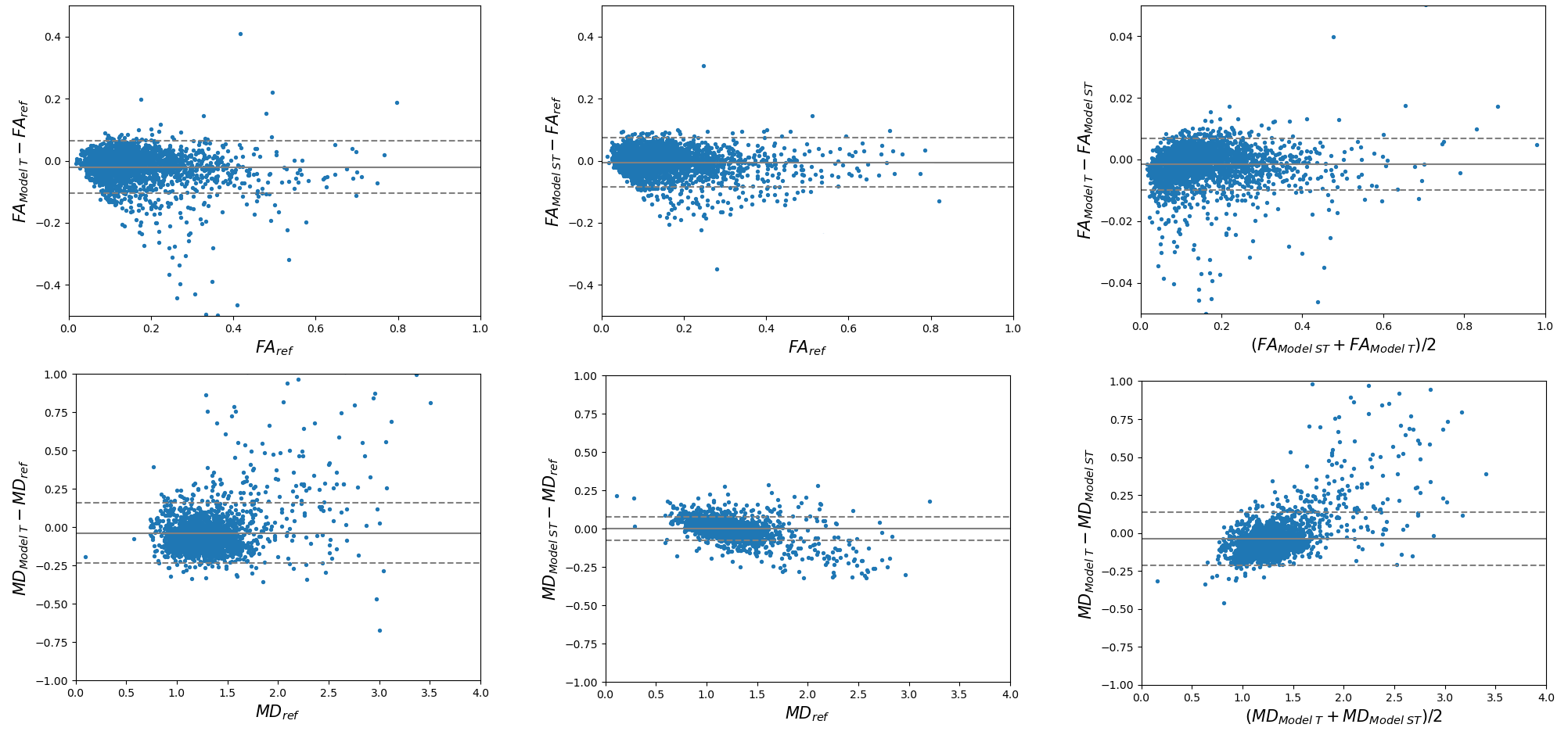}
\caption{Bland Altman plots for FA (top) and MD (bottom). Plots have been shown for Model T versus the reference (left column), Model ST versus the reference (middle column), and Model T versus Model ST (right column). To simplify the presentation we have multiplied values of MD by 1000 and they are in units of $mm^2s^{-1}$.}
\label{bland_altman}
\end{figure*}

Figure \ref{bland_altman} shows example Bland-Altman plots for the FA and MD errors for Model S and Model ST. We have shown plots for Model T versus the reference (left column), Model ST versus the reference (middle column), and Model T versus Model ST (right column). In the plots for Model T versus Model ST, because none of them can be considered the reference, we have used the average of the two models as the horizontal axis as suggested in \cite{bland1986statistical,giavarina2015understanding}. These plots show that for both FA and MD, both models tend to have larger errors for larger values of FA and MD. However, for both FA and MD, Model ST shows fewer large errors. For both FA and MD, the bias (i.e., the mean error compared with the reference) of Model ST is smaller than the bias of Model S. The plots in the right column in this figure also show that the ``corrections'' made by Model ST over Model S are over all values of FA and MD.

\begin{figure*}[!htb]
\centering
\includegraphics[width=\textwidth]{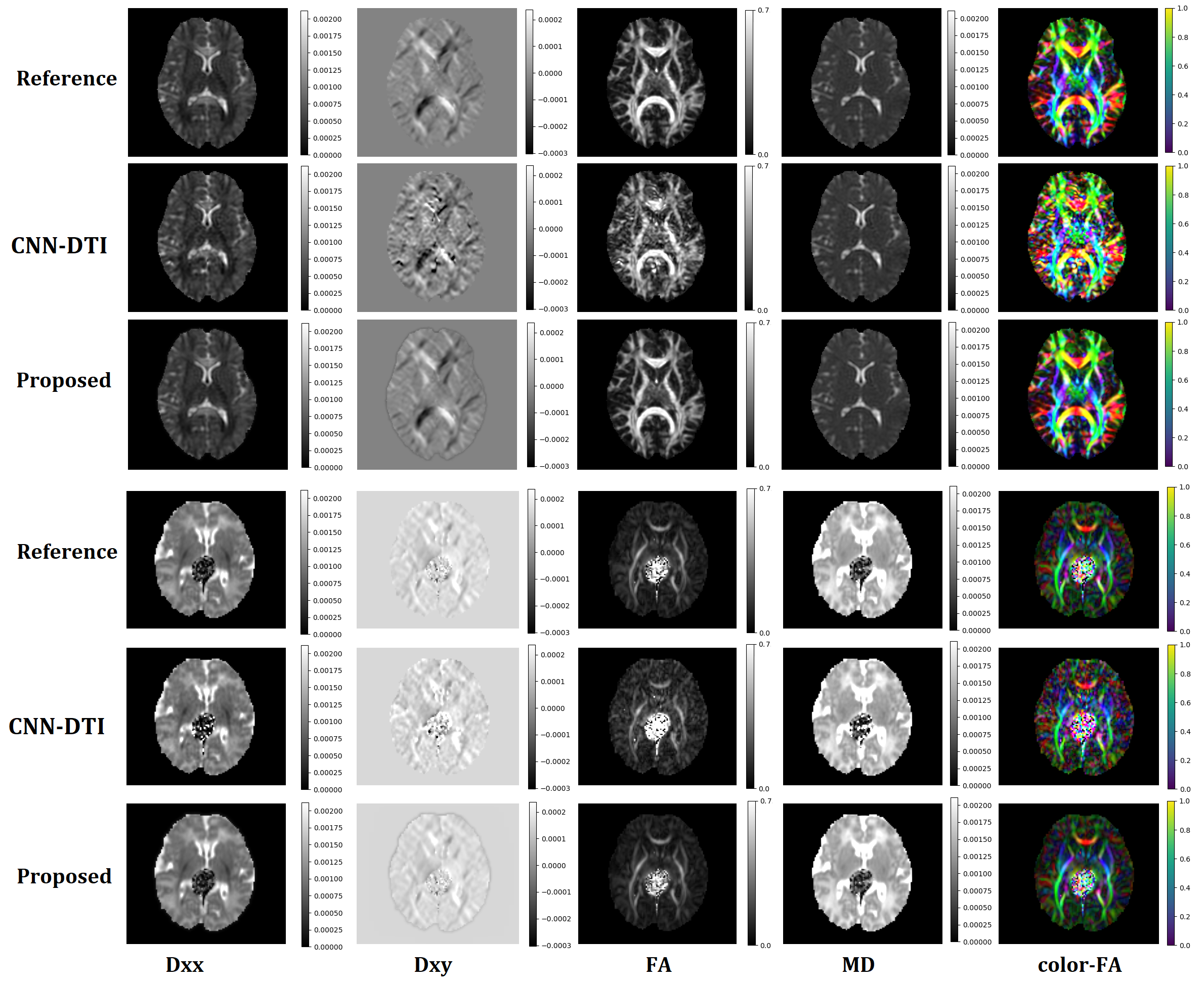}
\caption{Example tensor images and tensor-derived parameters obtained with CNN-DTI and the proposed method on a subject from the PING dataset (top) and a VOGM patient (bottom). The PING subject was 14 years old at scan time. The VOGM patient was 2 months old at scan time. The location of malformation in the brain of the VOGM patient is clearly visible in the reference image. For each subject, we have shown two tensor channels ($D_{xx}$ and $D_{xy}$), FA, MD, and color-FA.} \label{figure_ping_vogm}
\end{figure*}

Table \ref{table:results_ping_vogm} and Figure \ref{figure_ping_vogm} show the results of further evaluations of our method on 20 test subjects from the PING dataset and 7 VOGM patients. Each of the scans in these two datasets included 30 measurements in the $b=1000$ shell. For each scan in these datasets, we used all 30 measurements to reconstruct the reference image. We then selected six of the $b=1000$ measurements (and one $b=0$ measurement for normalization) for reconstruction with the proposed method and the three competing methods in the same way as described above for the experiments with the dHCP dataset. The VOGM patients' age range was 0-3 years and the PING subjects' age range was 3-20 years. Because of higher myelination and better data quality, competing methods achieved more accurate estimations on these two datasets than on the dHCP dataset. Moreover, compared with the experiments on the dHCP dataset above, in these experiments $D_{\text{ref}}$ is more biased towards $D_{\text{CWLLS}}$ and, to some extent, $D_{\text{CNLS}}$. This is because $D_{\text{CWLLS}}$ and $D_{\text{CNLS}}$ are reconstructed using 6 out of the 30 measurements used to reconstruct $D_{\text{ref}}$ and they are all reconstructed using least-square principles. Nonetheless, our proposed method achieved significantly lower errors on both of these additional datasets, as indicated in Table \ref{table:results_ping_vogm}. Compared with the competing methods, our proposed method reduced the estimation error in MD by factors of $1.9-3.2$, error in FA by factors of $1.4-1.55$, and error in the orientation of the major eigen-vector by factors of $1.38-1.58$. The superiority of the proposed method can be visually observed in the examples shown in Figure \ref{figure_ping_vogm}. Due to space limits, we have shown the results of our proposed method and CNN-DTI, which was slightly more accurate than CWLLS and CNLS. The differences between our method and CNN-DTI is especially more clear in the color-FA images that encode both the degree of anisotropy and the direction of the major eigen-vector in the same image.

Our study has some limitations that need to be mentioned. First, we only considered the diffusion tensor reconstruction from six measurements. The goal of this paper was to demonstrate the potential gains of exploiting the spatial correlations and the capability of attention-based neural networks to learn these correlations. Therefore, to enhance the focus of the study we used a fixed measurement scheme as input to our network. Although we anticipate that the use of these neural network models should be useful for other DW-MRI models and other measurement schemes, those could be investigated in future works. Moreover, although we have evaluated our method on three different and independent datasets, an application-specific evaluation may be warranted if our method is to be used for a specific clinical application. For example, it is not clear how our method (using six measurements) would compare with the reference image (using much larger measurement sets) if the reconstructions are to be used for detecting subtle changes in the brain micro-structure due to lesions or other pathologies. Therefore, further evaluation of our proposed method for specific clinical patient populations may be useful.

\section{Conclusions}

Diffusion tensor imaging is increasingly being used to study brain development and degeneration. However, the same least squares-based estimation methods \cite{koay2006unifying} have been commonly used in the past two decades. Standard estimation methods, which are based on linear or non-linear fitting of measurements on a per-voxel basis, can be highly sub-optimal. The main intuition of our work is that state of the art deep neural networks for modeling signal sequences can be adapted to develop accurate DTI estimation methods. In particular, the transformer models that have been used in this work are capable of learning very complicated correlations between signals in a sequence. Using this model, we developed a method that was able to learn spatial correlations between diffusion signals and tensor values in neighboring voxels for accurate tensor estimation. On the challenging neonatal DW-MRI scans from the dHCP dataset, our method reduced the estimation error, compared with three competing methods, by factors of 1.5-9.8. The estimations of the proposed method were close to the reference estimations that were obtained using 88 measurements. Our method also led to superior tractography and connectivity analysis. Furthermore, our method showed highly accurate and robust estimation on two additional datasets. These observations demonstrate the significant potential of the proposed method for improving the accuracy of DTI estimation, especially for challenging cohorts such as neonates and infants. Therefore, our method can facilitate DTI studies to detect subtle changes in brain, while also reducing the scan time.

\section*{Conflict of interest statement}

Authors do not have any conflicts of interests to disclose.

\bibliographystyle{ieeetr}
\bibliography{davoodreferences}

\end{document}